\documentstyle[11pt]{article}

\let\a=\alpha \let\b=\beta

 \def\bd{\begin{document}} \def\ed{\end{document}}
\def\ds{\documentstyle} \let\fr=\frac \let\bl=\bigl \let\br=\bigr
\let\Br=\Bigr \let\Bl=\Bigl 
\let\bm=\bibitem
\let\na=\nabla
\let\pa=\partial \let\ov=\overline 
\newcommand{\be}{\begin{equation}} 
\newcommand{\ee}{\end{equation}} 
\def\ba{\begin{array}}
\def\ea{\end{array}}
\def\ft#1#2{{\textstyle{{\scriptstyle #1}\over {\scriptstyle #2}}}}
\def\fft#1#2{{#1 \over #2}}
\def\del{\partial}
\def\vp{\varphi}
\def\sst#1{{\scriptscriptstyle #1}}
\def\oneone{\rlap 1\mkern4mu{\rm l}}
\def\td{\tilde}
\def\wtd{\widetilde}
\newcommand{\ho}[1]{$\, ^{#1}$}
\newcommand{\hoch}[1]{$\, ^{#1}$}
\newcommand{\bea}{\begin{eqnarray}} 
\newcommand{\eea}{\end{eqnarray}} 
\newcommand{\ra}{\rightarrow}
\newcommand{\lra}{\longrightarrow}
\newcommand{\Lra}{\Leftrightarrow}
\newcommand{\ap}{\alpha^\prime}
\newcommand{\bp}{\tilde \beta^\prime}
\newcommand{\tr}{{\rm tr} }
\newcommand{\Tr}{{\rm Tr} } 
\newcommand{\NP}{Nucl. Phys. }
\newcommand{\tamphys}{\it Center for Theoretical Physics\\
Texas A\&M University, College Station, Texas 77843}
\newcommand{\auth}{N. Khviengia, Z. Khviengia, H. L\"u\hoch{\dagger},
 and C.N. Pope\hoch{\dagger}}

\begin{document}

\hfill{CTP TAMU-19/96}

\hfill{hep-th/9605077}

\vspace{20pt}

\begin{center}
{ \large {\bf Intersecting M-branes and Bound States}}

\vspace{30pt}

\auth

\vspace{15pt}

{\tamphys}

\vspace{40pt}

\underline{ABSTRACT}
\end{center}

In this paper, we construct multi-scalar, multi-center $p$-brane solutions
in toroidally compactified M-theory.  We use these solutions to show that
all supersymmetric $p$-branes can be viewed as bound states of certain basic
building blocks, namely $p$-branes that preserve $1/2$ of the supersymmetry.
We also explore the M-theory interpretation of $p$-branes in lower
dimensions.  We show that all the supersymmetric $p$-branes can be viewed as
intersections of M-branes or boosted M-branes in $D=11$. 

{\vfill\leftline{}\vfill
\vskip	10pt
\footnoterule
{\footnotesize
     \hoch{\dagger}	Research supported in part by DOE 
Grant DE-FG05-91-ER40633 \vskip	-12pt} 	
}

\pagebreak
\setcounter{page}{1}

\section{Introduction}

     The construction and classification of $p$-brane solitons has been 
increasingly recognised to be important for gaining insights into the 
non-perturbative structures underlying string theory and M-theory. In 
particular, with the developing interest in eleven-dimensional M-theory, it 
is of interest to try to classify and interpret all its the extended-object 
solutions.  In $D=11$ itself, there are only two isotropic solutions, namely 
the elementary membrane \cite{dust} and the solitonic 5-brane \cite{g}.  As one 
descends through the dimensions to obtain lower-dimensional supergravities 
by dimensional reduction, a plethora of isotropic $p$-brane solutions arises. 
A fairly complete classification of these in toroidally-compactified
M-theory was given in \cite{lpsol}. 

     There are several possible approaches that can be taken in order to try
to organise these lower-dimensional $p$-branes into a more coherent picture.
One interesting idea, which was applied to black-hole solutions in $D=4$,
was to describe the various inequivalent supersymmetric black holes as bound
states of a fundamental ``building block,'' namely the single-charge
$a=\sqrt 3$ black hole that preserves $1/2$ of the supersymmetry
\cite{dr,sen,dlr,ct,r}.  Exploiting the fact that extremal black holes
satisfy a ``no-force'' condition, which allows the construction of
multi-center black hole solutions, it was shown in \cite{r} that all the
other supersymmetric black holes in $D=4$, corresponding to configurations
where multiple charges are non-zero, could be obtained as bound states in
the multi-center $a=\sqrt3$ solutions. 

      One of the purposes of the present paper is to extend the bound-state
discussion to a general setting.  Specifically, we shall show that all
supersymmetric $p$-brane solutions in M-theory can be described in terms of
certain fundamental $p$-branes that all preserve $1/2$ of the supersymmetry.
 To do this, we first construct general classes of multi-center
supersymmetric $p$-branes.  In order to encompass as broad a class of
solutions as possible, we consider multi-scalar solutions, in which
generically there are $N$ non-vanishing dilatonic scalar fields and $N$
independent charges carried by different field strengths, and show how to
construct multi-center generalisations in which each of the the
multi-centers for each of the $N$ charges can be located independently at
different positions in the transverse space.  In particular, when the locations
of all the charges are coincident the solution reduces to an isotropic
multi-scalar soliton \cite{lpmulti}.  Furthermore, if in addition all the
charges are set equal, they become single-scalar solutions.   There are also
a few examples of extremal $p$-branes that are non-supersymmetric, but which
are related to supersymmetric solutions simply by reversing the signs of
certain charges \cite{lpmulti,ko}.  These can also be viewed as bound states
of the same basic building blocks.  It should also be emphasised, however,
that there are many more isotropic single-scalar $p$-brane solutions
\cite{lpsol} that, although extremal, are neither supersymmetric nor related
by sign reversals to supersymmetric solutions.  Although they can be
generalised to multi-center solutions \cite{lps}, they do not admit suitable
multi-scalar generalisations that would enable them to be viewed as bound
states. 

     Another approach to classifying the lower-dimensional $p$-brane 
solutions in M-theory is to retrace the steps in the Kaluza-Klein 
dimensional reduction of the supergravity theory, and thereby look at the 
oxidations of the $p$-branes back to $D=11$.  The forms of the resulting 
configurations depend crucially on the details of which particular field 
strengths in the lower dimension are involved in the construction of the 
$p$-brane solution.  In certain simple cases, the oxidised solution simply 
becomes the isotropic membrane or 5-brane of the eleven-dimensional theory.
More commonly, however, the oxidised solution exhibits less than maximal 
isotropy either in the world-volume dimensions of the resulting extended 
object, or in the transverse space, or both.  Some of these configurations 
can be interpreted as describing intersections of membranes or 5-branes in 
$D=11$ \cite{pt,t,kt2,gkt,bl,lps}. If the field strengths participating in
the lower-dimensional solution include those that come from the $D=11$
vielbein under Kaluza-Klein dimensional reduction, this anisotropy will
include off-diagonal terms in the $D=11$ metric describing some kind of
twist, or non-trivial fibration, in the spacetime.  In some cases, these
off-diagonal terms have been interpreted as describing momentum flow
\cite{sv} along the spatial dimensions of the extended object.  In this
paper, we shall also explore some of these oxidation pathways in more
detail, and show how some of the lower-dimensional $p$-branes may be
interpreted back in $D=11$.

\section{Multi-scalar multi-center $p$-branes}

     Let us begin by constructing the multi-scalar, multi-center solutions.  
The relevant bosonic part of the supergravity Lagrangian, describing $N$
dilatonic scalar fields and $N$ $n$-index antisymmetric tensor field
strengths is given by \cite{lpmulti}
\begin{equation}
{\cal L}=eR- \frac{1}{2} e(\partial\vec \phi)^2-\fft{1}{2n!}e
\sum_{\alpha=1}^Ne^{-\vec a_\a \cdot \vec \phi}(F^{\alpha})^2\ .
\label{multilag}
\end{equation}
The metric ansatz is given by
\begin{equation}
ds^2=e^{2A} dx^{\mu}dx^{\nu}\eta_{\mu\nu}+e^{2B}dy^mdy^m
\ ,\label{metricform}
\end{equation}
where $x^\mu$ $(\mu=0,...,d-1)$ are the coordinates of the $(d-1)$-brane world
volume, and $y^m$ are the coordinates of the $(D-d)$-dimensional transverse
space. The functions $A$ and $B$, as well as the dilatonic scalars, depend
on the transverse coordinates $y^m$ only. For each $n$-index field strength
$F^\a$, one can construct an elementary $(n-2)$-brane carrying electric
charges with world volume dimension $d=n-1$, or a solitonic $(D-n-2)$-brane
carrying magnetic charges with $d=D-n-1$. In this paper, we shall consider
cases for $N\ge2$ where the field strengths all carry either electric or
magnetic charges.  For the case $N=1$, we shall also consider the dyonic
string in $D=6$. 

     We shall first consider the case where the field strengths carry 
electric charges, for which
\be
F^\a_{m\mu_1\cdots\mu_{n-1}} = \epsilon_{\mu_1\cdots\mu_{n-1}} \del_m 
e^{C_\a}\ ,\label{elefans}
\ee
where the functions $C_\a$ depend only on the transverse-space
coordinates $y^m$.  It is straightforward to obtain the equations of motion
following from the Lagrangian (\ref{multilag}), and to substitute the
ans\"atze for the field strength and the metric.  As in the case of the
isotropic multi-scalar solutions, one may obtain simple solutions by making
a further ansatz, namely $dA+ \td dB=0$.  Here, however, we do not assume
isotropicity in the transverse space.  In order to obtain multi-center
solutions, we first show that the resulting equations of motion can be cast
into a linear form.  The remaining equations of motion are given by 
\bea
&&\del_m\del_m \varphi_\a = \ft12 \sum_{\b} M_{\a\beta}\, S_m^\beta
S_m^\beta \ ,\label{dilatoneq} \\
&&\del_m\del_m A = \fft{\td d}{2(D-2)} \sum_\a S^\a_m S^\a_m\ , 
\label{aeq}\\
&&d(D-2) \del_m A\,\del_n A + \ft12 \td d \sum_{\a,\beta} (\del_m\varphi_\a)
(\del_n\varphi_\beta)=\ft12\td d\sum_\a S_m^\a S_n^\a\ ,\label{firstorder}\\
&& \del_m\del_m C_\a + \del_m C_\a (\del_m C_\a - 2 d\del_m A-\del_m 
\varphi_\a) = 0\ ,\label{ceq}
\eea
where we have made following definitions
\bea
\varphi_\a &=& \vec a_\a \cdot \vec \phi\ ,\qquad
M_{\a\beta} = \vec a_\a \cdot \vec a_\beta\ ,\nonumber\\
S^\a_m &=& e^{-\ft12 \varphi - d A}\, \del_m e^{C_\a} \ .\label{variousdef}
\eea
It follows from equations (\ref{dilatoneq}) and (\ref{aeq}) that it is 
natural to solve for $A$ by taking
\be
A = \fft{\td d}{D-2} \sum_{\a,\beta} (M^{-1})_{\a\beta} \,\varphi_\a\ .
\label{aphi}
\ee

    In the case of isotropic $p$-branes, the supersymmetric multi-scalar 
$p$-brane solutions arise if the dot products $M_{\a\beta}$ of 
the dilaton vectors $\vec a_\a$ satisfy \cite{lpmulti}
\be
M_{\a\beta} = 4 \delta_{\a\beta} - \fft{2d\td d}{D-2}\ ,\label{mmatrix}
\ee
where $\td d= D-d-2$.  It was shown in \cite{lpx} that the equations of
motion describing supersymmetric multi-scalar $p$-brane solutions can be
reduced to a set of $N$ Liouville equations with vanishing Hamiltonian. 
Since these solutions are supersymmetric and hence extremal, the no-force
condition is satisfied, and we can generalise them to multi-center
solutions.  On the other hand, for the dot products that do not satisfy
(\ref{mmatrix}), the equations of motion can be reduced to Toda-like
equations, whose solutions are intrinsically non-extremal \cite{lpx}.  We do
not expect that such solutions admit multi-scalar generalisations. 

     The inverse of the matrix $M_{\a\beta}$ (\ref{mmatrix}) is given by 
\be
M^{-1}_{\a\beta} = \ft14 \delta_{\a\beta} + \fft{d\td d}{4(2D-4 - d\td d N)}
\ .
\ee
For $N\ge 2$, it is singular in two cases, namely for $N=3$ black holes or 
strings in $D=5$, and for $N=4$ black holes in $D=4$.  We shall proceed for
now by assuming $M_{\a\beta}$ is invertible, and return to these singular
cases later, when we shall see that the general solution also solves the
equations in these cases.  Since $M_{\a\beta}$ is given by (\ref{mmatrix}),
it follows from (\ref{dilatoneq}) and (\ref{aeq}) that the equations
diagonalise with respect to the index $\a$, {\it i.e.}\ they take the form
$\del_m\del_m (\ft12 \varphi_\a + dA) =S^\a_mS^\a_m$ for each value of
$\alpha$.  These equations admit solutions for which $S^\a_m = \del_m
(\ft12\varphi_\a +d A)$, implying 
that 
\be
\del_m\del_m e^{-\ft12\varphi_\a - dA} =0\ ,\label{lineareq}
\ee
and $C_\a = \ft12\varphi_\a + dA$. It can easily be verified that the
remaining two equations (\ref{firstorder}) and (\ref{ceq}) are now also
satisfied.  Thus we have reduced the original equations of motion to a set of
$N$ linear equations (\ref{lineareq}), which admit multi-centered solutions 
\be 
e^{-\ft12\varphi_\a - d A}\equiv H_\a = 1 + \sum_i \fft{k^\a_i}{|\vec y - \vec
y^\a_i|^{\td d}}\ ,\label{gensol}
\ee
where $k_i^\a$ is the $i$'th charge for the $\a$'th field strength, located
at $\vec y^\a_i$ in the transverse space.  Note that the charges are
independent in number, size, and location, for each harmonic function
$H_\a$.  The metrics of the elementary multi-center multi-scalar $p$-brane
solutions are therefore given by 
\bea
ds^2 &=& e^{2A} dx^\mu dx^\nu \eta_{\mu\nu} + e^{2B} dy^mdy^m\ ,\nonumber\\
e^{2A} &=& \prod_{\a=1}^{N} \Big( 1 + \sum_i \fft{k^\a_i}{|\vec y - \vec
y^\a_i|^{\td d}} \Big)^{-\ft{\td d}{D-2}}\ ,\nonumber\\
e^{2B} &=& \prod_{\a=1}^{N} \Big( 1 + \sum_i
\fft{k^\a_i}{|\vec y - \vec y^\a_i|^{\td d}}\Big)^{\ft{d}{D-2}}\ .
\label{metricsol}
\eea

      The above discussion assumed that the matrix $M_{\a\beta}$ was
invertible.  The solutions contain $N$ dilatonic scalar fields, which is
consistent with having independent charges for the $N$ field strengths. If
the matrix $M_{\a\beta}$ is singular, which can arise in the two cases
$N=3$, $D=5$ and $N=4$, $D=4$ that we discussed above, the generic form of
the metric (\ref{metricsol}) continues to solve the equations of motion.
However, in this case it follows from (\ref{aphi}) and (\ref{gensol}) that
the dilatonic scalar fields are no longer independent, but satisfy
$\sum_{\a} \varphi_\a = 0$.  Thus the solutions in these two cases contain
$N$ independent charges, but only $(N-1)$ dilatonic scalar fields. 

     Having obtained the explicit elementary solutions, it is now 
straightforward to construct the solitonic solutions.  The solitonic ansatz for 
the field strength is given by
\be
F^\a_{m_1\cdots m_n} = -\epsilon_{m_1\cdots m_n p} \del_p 
\sum_{i} \fft{k^\a_i}{|\vec y - \vec y^\a_i|^{\td d}}\ .
\ee
The solitonic solutions can be obtained from the corresponding elementary
ones by making the replacement $\varphi_\a \rightarrow -\varphi_\a$.  It is
worth remarking that the terms ``elementary'' and ``solitonic'' are being
used with respect to the field strengths given in the Lagrangian
(\ref{multilag}).  In fact in $D=2n$ dimensions, some of the field strengths
in the Lagrangian (\ref{multilag}) can be the duals of the original field
strengths of the same rank in the supergravity theory.  Thus in terms of the
original fields, the solutions may carry both electric and magnetic charges.
Such solutions were called dyonic $p$-branes of the first type in
\cite{lpsol}. The solutions (\ref{metricsol}) also include multi-center
dyonic strings of the second type in $D=6$, where a single 3-form field
strength carries both electric and magnetic charges.  Note that the electric
and magnetic charges $k^\a_i$ in the multi-center harmonic functions can be
located at different places. 

\section{Extremal $p$-branes as bound states} 

     In the previous section, we obtained the general supersymmetric
multi-scalar multi-center $p$-brane solutions.  When all the charges are
equal and located in the same point in the transverse space, the solutions
reduce to isotropic single-scalar solutions for the Lagrangian 
\be
e^{-1}{\cal L} = R - \ft12 (\del\phi)^2 -\fft{1}{2n!} e^{-a\phi} F^2
\ ,\label{sslag} 
\ee 
where the constant $a$ can be parameterised as $a^2 = \Delta -2d\td d/(D-2)$ 
with $\Delta = 4/N$.  The supersymmetric isotropic single-scalar $p$-branes
in toroidally-compactified M-theory in all dimensions were classified in
\cite{lpsol}. It follows from our results in section 2 that the
multiply-charged solutions can be regarded as bound states of singly-charged
$p$-branes.  Consider a $p$-brane in which each field strength $F^\a$
carries a single charge $k^\a$ located at $\vec y^\a$, for which the metric
is given by 
\bea
ds^2 &=& e^{2A} dx^\mu dx^\nu \eta_{\mu\nu} + e^{2B} dy^mdy^m\ ,\nonumber\\
e^{2A} &=& \prod_{\a=1}^{N} \Big( 1 + \fft{k_\a}{|\vec y - \vec
y_\a|^{\td d}} \Big)^{-\ft{\td d}{D-2}}\ ,\nonumber\\
e^{2B} &=& \prod_{\a=1}^{N} \Big( 1 +
\fft{k_\a}{|\vec y - \vec y_\a|^{\td d}}\Big)^{\ft{d}{D-2}}\ .
\label{boundstates}
\eea
The metric approaches that of a singly-charged $p$-brane in the vicinity of
each of the locations $\vec y^\a$.  If these locations are chosen to be
coincident, the solution reduces to a multi-scalar $p$-brane solution with
independent charge parameters for the $N$ field strengths. If in addition
the charges are chosen to be equal, this solution then reduces to a
single-scalar solution with $\Delta = 4/N$.  Thus these multiply-charged
single-scalar or multi-scalar $p$-brane solutions can be viewed as bound
states of singly-charged $p$-branes.  In other words, all single-scalar
supersymmetric $p$-branes with $\Delta =4/N = 2, 4/3, 1, \ldots$, as well as
all their multi-scalar generalisations, can be described as bound states of
basic $\Delta=4$ single-charge building blocks (each of which by itself
would preserve $1/2$ of the supersymmetry).  In addition, there are certain
non-supersymmetric $p$-branes with $\Delta=4/N$ for $N\ge 4$, which are
related to supersymmetric ones simply by reversing the signs of certain 
charges \cite{lpmulti,ko}.  These too can be described as bound
states of the same $\Delta=4$ building blocks.  (Note however that the 
non-supersymmetric extremal $p$-branes with $\Delta \ne 4/N$ do not appear 
to admit a bound-state interpretation, since the necessary multi-scalar 
generalisations do not exist.) Of course, since all of the above $p$-brane
examples are constructed from configurations that obey a no-force condition,
they are bound states with zero binding energy. 

    It is worth remarking that the solutions (\ref{boundstates}) also
include the multi-center dyonic string in $D=6$, where the electric charge
and magnetic charges are located at different points in the transverse
space.  If the two location coincide, it becomes the standard isotropic
dyonic string. Thus the isotropic dyonic string \cite{dfkr}, including the
self-dual \cite{dl} and quasi-anti-self-dual \cite{lpsol,dlp2} limits, is a
bound state of the $\Delta=4$ electric string and the $\Delta=4$ magnetic
string.  

    All the $\Delta=4$ building blocks are singular-dilaton solutions, in
the sense that the dilaton field diverges at the horizon.\footnote{There do
in fact exist some $\Delta=4$ solutions that do not involve any dilaton, and
hence avoid this kind of singular behaviour, namely the M-branes (membrane 
\cite{dust} and 5-brane \cite{g}) in $D=11$, and the self-dual 3-brane in
$D=10$ type IIB supergravity \cite{hs,dl2}. However, these solutions cannot
be used to construct bound states, because, obviously, they do not admit the
necessary multi-scalar generalisations, and hence we shall not regard them
as building blocks for bound states.}   These divergences indicate that the
classical tree-level approximation is insufficient, and that string and
world-sheet loop corrections can be expected to modify the solutions
significantly.   It is interesting, however, that some of the bound states
built from these singular-dilaton building blocks are nevertheless
regular-dilaton solutions. Four such cases are known, namely the dyonic
string (with two charges) in $D=6$, the 3-charge black hole in $D=5$ (or its
string dual), the 4-charge black hole in $D=4$, and a special dyonic black
hole (also with four charges) \cite{lpsol} in $D=4$. In these solutions,
there are a total of $(N-1)$ non-vanishing dilatonic scalar fields, where
$N$ is number of the charges that the solution carries. These scalar fields
are regular at the horizon, and throughout the whole spacetime.  If the
charges are set equal, the dilatonic scalar field decouples, and the
solutions reduce to the self-dual string with $\Delta=2$ in $D=6$, the
Reissner-Nordstr\o{m} black holes with $\Delta=4/3$ or $\Delta=1$ in $D=5$
or $D=4$ respectively, and the special dyonic black hole with $\Delta=1$ in
$D=4$ \cite{lpsol}. These solutions are non-dilatonic, and are analogous to
M-branes in $D=11$ or the self-dual 3-brane in $D=10$. 

      Since these four bound states have dilatons that are regular
throughout the spacetime, they may not be appreciably modified by quantum
corrections. Thus the string theory may provide a microscopic interpretation
of the entropy for these solutions.  In particular, the
Reissner-Nordstr\o{m} black holes in $D=5$ and $D=4$ can be oxidised into a
boosted dyonic string in $D=6$ \cite{sv} and a boosted 3-charge string in
$D=5$ respectively.  Both these strings can be viewed as Dirichlet strings
carrying momentum, and hence their microscopic states can be counted.  Since
the entropy per unit $p$-volume is preserved under dimensional oxidation,
this provides a microscopic interpretation for the entropy of the
Reissner-Nordstr\o{m} black holes in $D=5$ \cite{sv} and $D=4$
\cite{jkm,sm}.  It is interesting to note that although the corresponding
building blocks, namely $\Delta=4$ black holes, have vanishing entropy,
these particular bound states, {\it i.e.}\ the Reissner-Nordstr\o{m} black
holes, nevertheless have non-vanishing entropy.  On the other hand, all the
other bound states, regular or singular, have vanishing entropy.  However,
it is worth remarking that all the regular bound states, if extrapolated to
the near-extremal regime, satisfy the ideal-gas relation $S\sim T^p$ between
their entropy and temperature, which is consistent with the D-brane picture
\cite{kt,lmpr}. This relation is not satisfied, however, by the singular
$p$-brane solutions.  It was argued in \cite{lmpr} that if the
singularities of the solutions were regulated by quantum effects, the
classically-singular $p$-branes would also satisfy this ideal-gas relation,
opening up the possibility of providing a microscopic interpretation of the
entropy for all $p$-branes. 

\section{Multi-center solitons and intersecting $p$-branes}

     Multi-center solutions provide a way of relating $p$-branes in 
different dimensions {\it via} the vertical dimension reduction procedure.
The procedure involves two stages \cite{lps}, namely an integration over a
continuum of charges of the multi-center solutions distributed uniformly
over lines, planes or hyperplanes, followed by an ordinary Kaluza-Klein
reduction over the resulting Killing directions.  It was shown in 
\cite{t,lps} that at the intermediate stage, the metric configuration in the
higher dimension can sometimes be interpreted as a special case of
intersecting $p$-branes.  For example, the membrane in $D=9$ can be viewed
as a multi-center membrane in $D=11$, with the charges lying on a plane in
the transverse space.  Remarkably, it was shown that this plane can be
elevated to the world-volume spatial dimensions of another elementary
membrane in a more general solution \cite{t}. This demonstrates that the
membrane in 9 dimensions can be viewed as two intersecting membranes in
$D=11$.  This procedure provides an 11-dimensional interpretation of a large
class of solutions, namely supersymmetric $p$-brane solutions whose charges
are carried by the field strengths derived from the 4-form in $D=11$.  If a
single M-brane is regarded as a special case of intersecting M-branes, 
these $p$-brane solitons in lower dimensions can be categorised in two
classes: 

\begin{description}

\item [1.] The Kaluza-Klein dimensional reduction of intersecting M-branes.
Examples are the $N$-charge black holes in $D=9,7,5$ and $3$ with $N=1,2,3$
and 4 respectively. These black holes can be viewed as $N$ intersecting
membranes in $D=11$, which preserve $2^{-N}$ of the supersymmetry
\cite{pt,t,gkt,lps}.  Another example is provided by the dyonic string in
$D=6$, which can be viewed as a membrane intersecting a 5-brane in $D=11$,
which was constructed in \cite{t}. 

\item  [2.] Vertical dimensional reduction of intersecting M-branes.  An
example is the membrane in $D=10$, which can be viewed as the multi-center 
membrane in $D=11$ with its charges distributed uniformly on the extra 
dimension.  Another example is the 2-charge black hole in $D=6$ that 
preserves $1/4$ of the supersymmetry.  It can be viewed as two intersecting
membranes in $D=11$, with the charges lying uniformly along a straight line.

\end{description}

    In the lower dimensions, there also exist many other supersymmetric
solitons whose charges are carried by the field strength derived from the
vielbein, rather than from the 4-form field strength in $D=11$.  These
solitons, together with the ones discussed above, form multiplets under the
U duality symmetry.  In this section, we shall discuss these solutions from 
the 11-dimensional point of view.  We shall show that can be viewed in
$D=11$ as boosted intersecting M-branes that carry momenta. 

       Let us first take black holes in $D=9$ as an example. There are in
total three 2-forms in $D=9$: ${\cal F}^{(1)}$ and ${\cal F}^{(2)}$
coming from the vielbein, and $F^{(12)}$ coming from the 4-form in $D=11$.
There are two two-scalar black holes, one using the 2-form field strength
$F^{(12)}$ and ${\cal F}^{(1)}$, the other using $F^{(12)}$ and ${\cal
F}^{(2)}$ \cite{lpmulti}.  We shall call them A-type and B-type black holes
respectively.  Both solutions preserve $1/4$ of the supersymmetry.  In fact
they form a doublet under the Weyl group $S_2$ of the U duality group
$SL(2,Z)$ in $D=9$ \cite{weyl}. The metrics for the two solutions are identical,
given by 
\be
ds_9^2 = -(H \wtd H)^{-\ft67} 
dt^2 + (H \wtd H)^{\ft17} dy^m dy^m\ ,\label{d9black}
\ee
where $H$ is an harmonic function in the 8-dimensional transverse space
associated with the field strength $F^{(12)}$, and $\wtd H$ is an
independent harmonic function associated with ${\cal F}^{(1)}$ or ${\cal
F}^{(2)}$ in the A-type or B-type black holes respectively.  We shall first
study them from the 10-dimensional perspective.  Since the 2-form field
strength ${\cal F}^{(1)}$ exists already in $D=10$, the two-scalar A-type
black hole, upon oxidation to 10 dimensions, gives rise to a string
intersecting a black hole, with metric given by 
\be
ds_{10}^2 = -H^{-\ft34} {\wtd H}^{-\ft78} dt^2 + H^{-\ft34} 
{\wtd H}^{\ft18} dz_2^2 
+ H^{\ft14} {\wtd H}^{\ft18} dy^m dy^m\ .\label{stringblack}
\ee
This interpretation can be seen by looking at the two limiting cases where 
either $H=1$, in which case (\ref{stringblack}) describes a multi-center black 
hole with its charges distributed uniformly along $z_2$, or $\wtd H=1$, in 
which case it describes an isotropic string.  Consideration of the generic
interpolating solutions, where $H$ and $\wtd H$ are both non-trivial, then
completes the interpretation.  Note that the string and the black hole share
a common time coordinate and a common transverse subspace. The two-scalar
B-type black hole, on the other hand, is quite different from the
10-dimensional point of view.  Its metric is given by 
\be 
ds_{10}^2 = - H^{-\ft34} {\wtd H}^{-1} dt^2 + H^{-\ft34} 
{\wtd H} (dz_2 + {\cal A}^{(2)})^2 + H^{\ft14} dy^m dy^m\ . 
\ee
This metric describes a boosted string carrying a momentum related to the
charge carried by ${\cal F}^{(2)}=d {\cal A}^{(2)}$.  The solution is
analogous to the boosted dyonic string in $D=6$ \cite{sv}, which is the
dimensional oxidation of the 3-charge black hole in $D=5$.

     We have seen that the two two-scalar black holes in $D=9$, which form a
doublet under the $D=9$ Weyl U duality, have a very different interpretation
in $D=10$.  The A-type black hole gives rise to an intersection of a string 
and a black hole in $D=10$, whilst the B-type gives rise to a boosted string. 
Such a difference reflects the fact that $D=10$ is not 
the fundamental dimension.  If we further oxidise the $D=10$ solutions to 
$D=11$, we obtain the two metrics
\bea
\hbox{A-type}:&&\!\!\! ds_{11}^2 = -H^{-\ft23} {\wtd H}^{-1} dt^2 +
H^{-\ft23} {\wtd H} (dz_1 + {\cal A}^{(1)})^2 + H^{-\ft23} dz^2_2 +
H^{\ft13} dy^mdy^m \ ,\nonumber\\
&& \label{eleven}\\
\hbox{B-type}:&& \!\!\!ds_{11}^2 = -H^{-\ft23} {\wtd H}^{-1} dt^2 +
H^{-\ft23}dz_1^2 + H^{-\ft23} {\wtd H} (dz_2 + {\cal A}^{(2)} )^2 +
H^{\ft13} dy^mdy^m \ .\nonumber
\eea
Thus both the A-type and B-type black holes in $D=9$ can be viewed as boosted
membranes in $D=11$, but carrying momentum in the two different spatial
world-volume directions $z_1$ and $z_2$.  The Weyl U duality in $D=9$ can be
reinterpreted as a duality symmetry $z_1 \leftrightarrow z_2$ in $D=11$.  It 
is interesting that the manifest U Weyl duality, which is present in $D=9$,
is somewhat obscured upon oxidation to $D=10$, but is restored in $D=11$.  
This can be taken as an indication that M-theory plays a more fundamental
role than string theory.

     The picture is more complicated in lower dimensions, where the U
duality group is larger.  For example, there are two kinds of 3-charge black
holes in $D=5$, one of which involves field strengths coming only from the
4-form in $D=11$, while the other involves a field strength coming from the
metric.  Upon oxidation the former becomes three intersecting membranes in
$D=11$, with the metric \cite{t}
\bea
ds^2_{11} &=& -(H_1 H_2 H_3)^{-\ft23} dt^2 + H_1^{-\ft23} (H_2 H_3)^{\ft13} 
(dz_1^2 + dz_2^2) + H_2^{-\ft23} (H_1 H_3)^{\ft13} 
(dz_3^2 + dz_4^2)\nonumber\\
&&+H_3^{-\ft23} (H_1 H_2)^{\ft13} 
(dz_5^2 + dz_6^2) +  (H_1 H_2 H_3)^{\ft13} 
(dy_1^2 +\cdots + dy_4^2)\ ,\label{3m}
\eea
where $H_1, H_2$ and $H_3$ are the harmonic functions associated with 
$F^{(12)}, F^{(34)}$ and $F^{(56)}$ respectively.  On the other hand, we 
find that the latter metric becomes 
\bea
ds^2_{11} &=& - H_1^{-\ft23} H_2^{-\ft13} H_3^{-1} dt^2 + 
H_1^{\ft13} H_2^{-\ft13} (dz_1^2 + \cdots + dz_4^2) +
H_1^{-\ft23} H_2^{\ft23} dz_5^2 \nonumber\\
&&+ H_1^{-\ft23} H_2^{-\ft13} H_3 (dz_6 + {\cal A}^{(6)})^2 +
H_1^{\ft13} H_2^{\ft23} (dy_1^2 + \cdots + dy_4^2)\ ,\label{1f1m}
\eea
where in this example, $H_1, H_2$ and $H_3$ are the harmonic functions
associated with $F^{(56)}$, $*F^{(5)}$ and ${\cal F}^{(6)}$ respectively.  
This describes a boosted membrane intersecting a 5-brane.  (Note that
Kaluza-Klein dimensional reduction of (\ref{1f1m}) to $D=6$ gives rise to
the boosted dyonic string constructed in \cite{sv}.)  The two 3-charge black 
holes in $D=5$ are related by U duality; in fact they belong to the same 
multiplet of the U Weyl group \cite{weyl}.  However, as we have seen, upon 
oxidation to $D=11$ the U duality is obscured, unlike the example of 
$D=9$ black holes that we discussed previously.  It is tempting to speculate 
that manifest U duality might be restored in a yet higher dimension.

      Having illustrated the procedure with two examples, we can easily
generalise these results to include all the supersymmetric $p$-branes in
lower dimensions, in which some of the charges are carried by the field
strengths coming from the vielbein. We shall categorise this type of
$p$-brane in the third class: 

\begin{description}

\item [3.] Kaluza-Klein or vertical dimensional reduction of boosted
intersecting M-branes. 

\end{description}

\section{Conclusions and discussions}

    In this paper, we constructed multi-scalar multi-center $p$-brane
solutions, which are generalisations of supersymmetric single-scalar
solutions with $\Delta = 4/N = 2, 4/3, 1,$ {\it etc}.  The solutions carry
$N$ independent sets of charges $k^\a_i$, ($1\le \a \le N$), which can be
located at different points $\vec y^\a_i$ in the transverse space.  In the
vicinity of each charge, the solution approaches a single-scalar solution
with $\Delta=4$.  If these locations coincide, the solution reduces to an
isotropic multi-scalar solution.  Furthermore, if the charges are set equal,
it becomes a single-scalar solution with $\Delta=4/N$.  The
multi-center solutions preserve the same fraction of the supersymmetry as
their isotropic limits where the centers coincide. Thus we conclude that all
supersymmetric isotropic single-scalar $p$-branes, and all their
multi-scalar generalisations, can be viewed as bound states of
supersymmetric $\Delta=4$ single-scalar building blocks.  In addition, there
are certain examples of $p$-branes with $\Delta=4/N$ for $N\ge 4$ that do
not have supersymmetry, owing to sign-reversals of some of the charges.
These too can be interpreted as bound states of the same basic building
blocks. All the building blocks are singular at the horizon, in that the
scalar fields, field strengths and curvature diverge there.   Interestingly,
some of their bound states are nevertheless regular.

     We also discussed the close relationship between vertical dimensional
reduction and the description of intersecting M-branes.   Consider a
multi-center $p$-brane in $D$ dimensions whose charges lie uniformly on an
$n$-dimensional hyperplane in the transverse space.  This is the first stage
in a vertical dimensional reduction to a $p$-brane in $(D-n)$ dimensions. On
the other hand, in certain cases the hyperplane can be promoted to the
status of being the spatial world-volume of a more general solution. Thus
vertical dimension reduction provides a way of interpreting certain
lower-dimensional $p$-branes as intersections of extended objects in higher
dimensions. We showed that all supersymmetric $p$-branes in lower dimensions
can be categorised by three classes. The first and second classes involve
$p$-branes whose charges are all carried by those field strengths that are
derived by dimensional reduction from the 4-form in $D=11$.  These solutions
can be viewed as either Kaluza-Klein or vertical dimensional reductions of
certain intersecting M-branes in $D=11$.  The third class of $p$-branes have
some charges that are carried by field strengths coming from the vielbein.
These solutions can be regarded as Kaluza-Klein or vertical dimensional
reductions of boosted intersecting M-branes. 

      Another interesting case that arises from the multi-scalar multi-center 
solutions is if each field strength carries charges distributed 
uniformly along {\it different} subspaces of the transverse space.  In 
particular, these subspaces can be non-overlapping.  For example, in the 
two-scalar black hole solution (\ref{d9black}), the electric charge $q$ carried
by $F^{(12)}$ can be chosen to lie uniformly in the $(y_1,y_2, y_3)$
hyperplane, while the electric charge $\td q$ carried by ${\cal F}^{(1)}$ or
${\cal F}^{(2)}$ can be chosen to lie uniformly in the $(y_4, \ldots, y_8)$
hyperplane. If $q$ is set to zero, the resulting metric configuration
describes the Kaluza-Klein oxidation of a black hole in six dimensions.  On
the other hand, if the charge $\td q$ is zero, the solution is the oxidation
of a black hole in four dimensions.  These solutions are reminiscent of the
previously-discussed intersecting $p$-branes, except that here the extended
objects share a common world-volume but have non-coincident transverse
spaces. 

\section*{Acknowledgement}

    We are grateful to M.J. Duff, J. Rahmfeld and K.S. Stelle for useful
discussions.

\end{document}